\documentclass[prl,twocolumn,aps,floatfix]{revtex4}
\usepackage{epsfig}
\usepackage{amsmath}
\usepackage{amssymb}
\usepackage{graphicx}
\usepackage{ulem}
\usepackage{multirow}
\usepackage[usenames]{color}
\usepackage[english]{babel}

\def\figx#1#2{\includegraphics[width=#1]{#2}}

\newcommand{\dg}{^{\dagger}}

\newlength{\bxwidth}

\newlength{\fight}
\fight= \columnwidth
\newcommand{\fg}[3]
{\begin{figure}[here]
\[ 
\figx{\fight}{#1}
\] 
\vspace*{-4mm}
\caption{\label{#2}
\small #3}
\end{figure} }
\def\gtappr{{{\lower4pt\hbox{$>$} } \atop \widetilde{ \ \ \ }}}
\def\ltappr{{{\lower4pt\hbox{$<$} } \atop \widetilde{ \ \ \ }}}

\newcommand{\pmat}[1]{\begin{pmatrix} #1 \end{pmatrix}}

\newcommand{\urs}{URu$_{2}$Si$_{2}$\ }
\newcommand{\ursp}{URu$_{2}$Si$_{2}$}

\newcommand \bea {\begin{eqnarray} }
\newcommand \eea {\end{eqnarray}}

\begin{document}
\title{Basal-Plane Nonlinear Susceptibility: A Direct Probe of the Single-Ion Physics in \urs}
\author{Rebecca Flint$^1$, Premala Chandra$^2$ and Piers Coleman$^{2,3}$}
\affiliation{$^1$ 
Department of Physics, Massachusetts Institute for Technology,
77 Massachusetts Avenue, Cambridge, MA 02139-4307}
\affiliation{$^2$ Center for Materials Theory, Department of 
Physics and Astronomy, Rutgers University, 
Piscataway, NJ 08854}
\affiliation{$^{3}$Department of Physics, Royal Holloway, University
of London, Egham, Surrey TW20 0EX, UK.}
\date{\today}
\begin{abstract}
The microscopic nature of the hidden order state in \urs is dependent
on the low-energy configurations of the uranium ions, and there
is currently no consensus on whether it is predominantly $5f^2$ or $5f^3$.  
Here we show that measurement of the
basal-plane nonlinear susceptibility can
resolve this issue; its sign at low-temperatures is a distinguishing factor.
We calculate the linear and nonlinear susceptibilities
for specific $5f^2$ and $5f^3$ crystal-field schemes that are
consistent with current experiment.
Because of its dual magnetic and orbital character, 
a $\Gamma_5$ magnetic non-Kramers
doublet ground-state of the U ion can be identified by 
$\chi_1^c(T) \propto \chi_3^\perp(T)$ where we have determined
the constant of proportionality for \urs.
\end{abstract}
\maketitle
%
%
%
%

\section{I. Motivation}

Despite tremendous experimental and theoretical efforts over the last
quarter century, the nature of the ordering (``hidden order'') at $T_0
\sim 17.5 K$ in the actinide heavy fermion \urs remains
unresolved\cite{Mydosh11}. Clear signatures of Fermi liquid
behavior\cite{Mydosh11,Palstra85} above $T_0$ and the sharp anomalies
in thermodynamic
properties\cite{Mydosh11,Palstra85,Miyako91,Ramirez92} at $T_0$
suggest a Fermi surface instability and thus itinerant
behavior\cite{Oppeneer10}.  However, the large entropy of condensation
($\frac{S}{N} \sim 0.3 R \ln 2$), the Curie-Weiss susceptibility at
room temperature\cite{Mydosh11,Palstra85} and the strong Ising
anisotropy observed in the magnetic responses\cite{Miyako91,Ramirez92}
all point to a localized origin of the key magnetic degrees of freedom
that contribute towards the formation of the hidden order state.

A key outstanding issue for \urs is
to determine whether its U ion ground-state configuration is predominantly
$5f^2$ or $5f^3$.  Specific heat, high-temperature susceptibility and
even photoemission measurements are unable to resolve this
issue\cite{Mydosh11}. Inelastic neutron scattering
experiments\cite{Broholm91,Park02} favor $5f^2$ whereas EELS
measurements\cite{Jeffries10} are consistent with $5f^3$.  
While techniques like inelastic
x-ray and neutron scattering can successfully resolve the crystal
field configurations of $4f$ ions like cerium and
praseodymium\cite{Severing10,Fulde85}, these techniques fail for $5f$
ions like uranium due to significantly broader crystal field levels.
Resolution of the ground state configuration of the
uranium ion in \urs would provide an important constraint on the theoretical
description of its hidden order.  

\fg{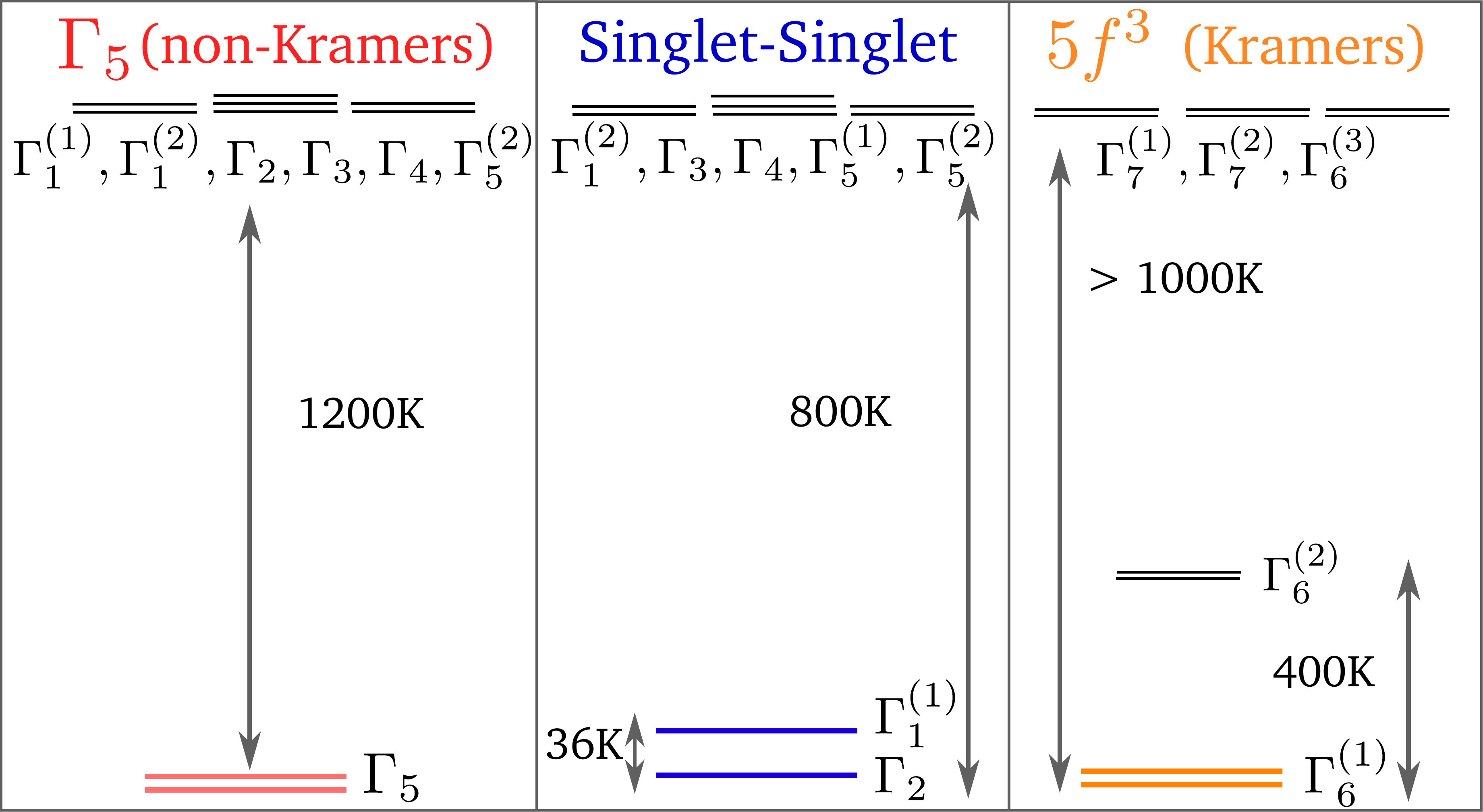}{fig1}
{{Three} possible crystal-field schemes for \ursp. (A) The $5f^2$
$\Gamma_5$ non-Kramers doublet ground-state, with all excited
crystal-field states degenerate at 1200 K. (B) Two finely-tuned
low-lying singlets scenario (with the two doublets found from
DMFT\cite{Haule09}), with all other states degenerate at 800 K to
reproduce the Van Vleck susceptibility.  (C) The $5f^3$ Kramers
doublet ground state, with a nearly Ising doublet ground state,
$\Gamma_6 = -.91 |\pm 7/2 \rangle + .42|\mp 1/2\rangle + .05|\mp
9/2\rangle$ 
chosen to best represent the measured Ising
susceptibility, with another, excited $\Gamma_6$ doublet at 400K and
all other excited doublets above 1000K.}

Here we present the nonlinear susceptibility ($\chi_3$)
as a probe of the nominal valence of the uranium ion in \ursp. In cubic UBe$_{13}$, 
$\chi_3$ provided strong evidence that the low-lying magnetic
excitations are predominantly dipolar ($5f^3$) rather than quadrupolar ($5f^2$) 
in character\cite{Ramirez94}.  We generalize this technique to the
tetragonal \urs system, where the sign of $\chi_3$ distinguishes between
the two nominal valences $5f^2$ and $5f^3$. 
The linear and the nonlinear
susceptibilities are calculated for specific $5f^2$ and $5f^3$ crystal-field schemes consistent
with recent experiment.
In addition  we discuss the dual nature of the magnetic non-Kramers doublet,
$\Gamma_5$, that is magnetic along the c-axis but quadrupolar in the
basal plane.  As both magnetic ($\chi_1^c$) and quadrupolar
($\chi_3^\perp$) susceptibilities originate from splitting the same
doublet, they will have the same temperature-dependence even outside
the single-ion regime.

\section{II. Time-Reversal Properties of $5f^2$ and $5f^3$ Configurations}

The two valence configurations $5f^2$ and $5f^3$ are clearly distinguished
by their properties under time-reversal. 
$5f^3$ is always a Kramers doublet; $5f^2$ can either involve low-lying singlets or, 
in the tetragonal environment
appropriate for \ursp, it can form a $\Gamma_{5}$ doublet formed
from two degenerate orbitals $\vert x\rangle $ and $\vert y\rangle $
that are crystal rotations of one another. 
In a tetragonal environment, these 
two states are degenerate, so one can construct a
magnetic basis $\vert \pm \rangle = \vert x \rangle \pm  i \vert y
\rangle $.
Such a ``non-Kramers doublet'' possesses a dual orbital and
magnetic character;
and it can be distinguished from a
Kramers doublet through its time-reversal properties. A non-Kramers doublet
time reverses according to 
\begin{equation}\label{}
\theta\vert  \pm \rangle = \vert
\pm \rangle ,
\end{equation}
whereas a Kramers doublet transforms with an additional and
all-important minus sign 
\begin{equation}\label{}
\theta \vert  \pm \rangle = \pm 
\vert \mp\rangle .
\end{equation} 
Consider next the case of two almost-degenerate low-lying singlets
$\vert a\rangle $ and $\vert b\rangle $
that are not related by symmetry, separated by a small crystal field
splitting $\Delta$, as in the scenarios considered in \cite{Haule09}
or \cite{Santini94}. At temperatures and
fields $B, \ T >> \Delta $, one can treat this almost-degenerate set of
singlets as an accidentally degenerate 
non-Kramers doublet, defined by the magnetic
states  $\vert \pm \rangle  = \vert  a\rangle \pm  i \vert  b \rangle
$.  In this way, the properties of a $5f^{3}$ and $5f^{2}$
configuration at temperatures and fields large compared to their
splitting can  be  considered as Kramers or non-Kramers
doublets respectively.

A distinctive property  of non-Kramers doublets is their 
Ising symmetry, a consequence of their time-reversal symmetry. 
The matrix
elements of a time-reversed operator are given by its complex conjugate
\cite{Sakurai94}
\begin{equation} 
\langle \alpha|A| \beta\rangle^{*}
=\langle \beta|A\dg | \alpha\rangle
=\langle \tilde{\alpha} \vert 
\Theta A\Theta^{-1}|\tilde{\beta}\rangle 
\label{sakurai}, 
\end{equation} 
where $\Theta |\alpha\rangle = |\tilde{\alpha}\rangle$ is the
time-reversal of state $\vert \alpha \rangle $. 
If we apply this relation to the off-diagonal matrix element 
$\langle - \vert J_{-}\vert +\rangle $, then 
\begin{equation} 
\langle -|J_{-}| +\rangle^{*}
=\langle +|J_{-}\dg | -\rangle
=\langle \widetilde{- }\vert
\Theta J_{-}\Theta^{-1}|\widetilde{+}\rangle 
\label{sakurai2}. 
\end{equation} 
Now 
on the left-hand side, since $J_{-}\dg = J_{+}$,  we
obtain $\langle - \vert J_{-}\vert + \rangle^{*} = \langle + \vert
J_{+}\vert - \rangle $. The action of 
time-reversal on $J_{-}=
J_{x}- iJ_{y}$ reverses the signs of the angular momentum operators,
and as an anti-unitary operator, takes the complex
conjugate of the coefficient $-i$; it thus follows that 
$\Theta
J_{-}\Theta^{-1}= - J_{+}$. The time-reversal of the non-Kramers
states is given by 
$\vert \widetilde{-}\rangle = \vert
+\rangle $ and $\vert \widetilde{+}\rangle = \vert
-\rangle $ so that on the right-hand side of (\ref{sakurai2}) , $\langle \widetilde{-}\vert \Theta
J_{-}\Theta^{-1}\vert \widetilde{+}\rangle = - \langle + \vert
J_{+}\vert -\rangle $. Comparing both sides, it follows that 
\begin{equation}\label{}
\langle-|J_-|+\rangle = - \langle+|J_-|-\rangle = 0,
\qquad (\hbox{non-Kramers doublet})
\end{equation}
with a
corresponding relation for $J_{+}$. It follows that the off-diagonal
matrix elements of $J_{\pm}$ vanish for a non-Kramers doublet, 
giving rise to a resulting
Ising anisotropy that is independent of crystal structure. 

By contrast, for a Kramers
doublet, the additional minus sign in the transformation of the
states removes the constraint on the off-diagonal matrix elements.
To see how this works for \ursp, suppose that the f-configuration is a
5f$^{3}$ Kramers doublet predominantly in an $\vert \pm 7/2\rangle $
state.  The presence of a tetragonal symmetry will add and subtract
units of $\pm 4 \hbar $, so that the crystal field ground-state will
take the form 
\begin{equation}\label{}
\vert \pm \rangle= a \vert \pm 7/2\rangle + b \vert \mp 1/2\rangle +c \vert \mp 9/2\rangle
\end{equation}
In this case, $\vert \langle - \vert J_{+}\vert + \rangle
\vert^{2} =  5b^{2} + 6 a c$, so 
perfect Ising anisotropy is only obtained when $5b^2 + 6 a c = 0$, i.e
if the tetragonal crystal fields that mix the two configurations are
fine-tuned to zero.

However if the ground-state is a non-Kramers
doublet of the form 
\begin{equation}\label{}
\vert \pm \rangle  = a \vert \pm 3 \rangle + b \vert \mp 1 \rangle 
\end{equation}
then an Ising anisotropy follows for arbitrary mixing between the
$\vert \pm 3\rangle $ state and the $\vert \mp 1\rangle $ state. 
Thus the observed Ising anisotropy in \urs either results from a
finely-tuned $5f^3$ state, or from a $5f^{2}$ state with a real or an
effective non-Kramers doublet ground-state.  The challenge is to
distinguish these scenarios. We now show that this can be done 
by measuring the nonlinear susceptibility to a transverse field
(perpendicular to the Ising axis).

Kramers and non-Kramers states differ in their response to   a
transverse field.  If we integrate out the high-lying crystal
field excitations in the presence of a transverse field in the
x-direction, the remaining effective Hamiltonian will contain
off-diagonal terms of the form 
\begin{equation}\label{}
{\cal H }_{eff} (B) = \pmat{0 &\Delta E (B)\cr\Delta E (B) & 0}
\end{equation}
Now for a Kramers doublet, applying (\ref{sakurai}), we obtain
\bea\label{}
\langle + \vert {\cal H}_{eff} (B)\vert  - \rangle &=& 
- \langle + \vert  \Theta {\cal H}_{eff} (B)\Theta^{-1}\vert -\rangle\cr
& = & - \langle  + \vert {\cal H}_{eff} (-B)\vert - \rangle 
\eea
so the off-diagonal matrix elements of ${\cal H}_{eff} (B)$ must be an
odd function of $B$, taking the form
\begin{equation}\label{kramers}
\Delta E_{K} (B) = g B + \frac{1}{3!}\gamma B^{3}.
\end{equation}
Here $\gamma\sim 1/\Delta_{VV}^{2}$ is a
consequence of third order perturbation theory, where $\Delta_{VV}$ is the gap to excited crystal field states, and $\gamma$ can have either sign.
Physically, this means that a Kramers doublet can develop dipole and
octupole components, but has no quadrupolar response. 
By contrast, for a non-Kramers doublet, applying (\ref{sakurai}), we
obtain
\begin{equation}\label{}
\langle + \vert {\cal H}_{eff} (B)\vert  - \rangle = 
 \langle + \vert  \Theta {\cal H}_{eff} (B)\Theta^{-1}\vert -\rangle
=  \langle  + \vert {\cal H}_{eff} (-B)\vert - \rangle 
\end{equation}
so the off-diagonal matrix elements of ${\cal H}$ are even in field,
where the leading order term is quadrupolar
\begin{equation}\label{nonkramers}
\Delta E_{NK} (B) = \frac{1}{2}q B^{2}.
\end{equation}
Here $q\sim 1/\Delta $, since this results from second-order
perturbation theory.

The nonlinear susceptibility, $\chi_3$, 
is defined as the cubic term in the magnetization
\begin{equation}
M = \chi_1B + (1/3!) \chi_3 B^3 +....
\end{equation}
in the direction of the applied field ($B$), so that
$\chi_{3}=\partial^{3}M/\partial B^{3}= -\partial^{4}F/\partial B^{4}
$.  If we take the 
the free energy $F (B) = - T \ln\biggl ( 2 \cosh \bigl [ \beta \Delta  E (B)\bigr]\biggr)$ and
do a high-temperature expansion, we find
\bea
F(B) & = & -T \ln \biggl \{ 2 \biggl [ 1 + \frac{1}{2} \biggl (\frac{\Delta E(B)}{T}\biggr)^2 + 
\frac{1}{4!} \biggl(\frac{\Delta E(B)}{T}\biggr)^4 \biggr ] \biggr \}\cr
& \sim & - \frac{1}{2} \frac{\bigl [\Delta E(B) \bigr ]^2}{T} + \frac{1}{12} \frac{\bigl [\Delta E(B) \bigr]^4}{T^3}.
\eea
Using expressions
(\ref{kramers}) and (\ref{nonkramers}), we obtain 
\begin{eqnarray}\label{l}
\chi_{3}^{NK} &=& \ \ \frac{3 q^{2}}{T},\qquad \qquad
(\hbox{Non-Kramers doublet})\cr\cr
\qquad \chi_{3}^{K} &= & \frac{4 g \gamma}{T} -\frac{3 g^2}{T^3} \qquad
(\hbox{Kramers doublet}).
\end{eqnarray}

We note that the quadrupolar
response of the non-Kramers doublet leads to a positive
$\chi^{NK}_{3}>0$, whereas its Kramers counterpart is negative at the lowest temperatures where the $1/T^3$ term dominates.  Thus the different time-reversal properties of a $5f^{3}$
Kramers doublet and a $5f^{2}$ non-Kramers doublet (or an effective
non-Kramers doublet composed of two low-lying singlets) can be
experimentally distinguished by the sign of
of the nonlinear susceptibility at low temperatures. 
However, it is important that the U remain in the single-ion regime.  
If $\gamma$ is positive there is a crossover temperature 
scale, $T_x \sim \sqrt{\frac{3g}{4\gamma}}$, 
where $\chi_3^\perp$ changes sign (but if $\gamma$ is negative then $\chi_3^\perp < 0$ for all T).  
Therefore, provided the single-ion 
regime continues to sufficiently low temperatures, one can identify the magnetic 
state of the U ion from the sign of the basal-plane nonlinear susceptibility.  
This experiment may require dilution studies to extend the single-ion regime down to sufficiently low temperatures.

\section{III. $\chi_3$ for Specific $5f^2$ and $5f^3$ Single-Ion Schemes}

In order to illustrate the above results in detail, we have calculated
the linear and nonlinear susceptibilities within three distinct
crystal-field schemes:
(i) non-Kramers doublet, (ii) a ``pseudo''-non-Kramers doublet given by
two closely spaced singlets and  (iii) a Kramers doublet.  These three
scenarios are chosen to reproduce the 
high temperature behavior: the
Ising anisotropy, moment and van Vleck susceptibilities.  We assume
that the maximum in $\chi_1^c$\cite{Palstra85} results from the Kondo
effect, and thus we are not considering ``three-singlet'' crystal
field schemes\cite{Santini94}.

The three crystal-field scenarios that we consider here (cf. Fig. 1) are:
\begin{itemize}
\item{(A) {\bf Non-Kramers doublet}: $5f^2$ (J=4) $\Gamma_5$:
\begin{equation}
\vert \Gamma_5 \pm \rangle = a \vert \pm 3\rangle + b \vert \mp 1\rangle,
\end{equation}
where $a,b$ are adjustable parameters.  For simplicity, all other crystal field states (five singlets and one doublet) have been placed at 1200K to reproduce the magnitude of the van Vleck susceptibility; however, the behavior of the linear and non-linear susceptibilities are relatively insensitive to the exact arrangement of the excited states.}
\item{(B) {\bf Finely Tuned Singlets
}: $5f^2$ (J = 4) with two low-lying singlets separated by a gap, $\Delta = 36K$.  We use the singlets, $\Gamma_2$ and $\Gamma_1$ proposed by Haule and Kotliar\cite{Haule09}:
\bea
\vert \Gamma_2 \rangle & = & \sqrt{\frac{1}{2}} \left( \vert 4\rangle -\vert-4\rangle\right)\cr
\vert \Gamma_1\rangle & = & \frac{\cos \phi}{\sqrt{2}}\left( \vert 4\rangle +\vert-4\rangle\right) -\sin \phi \vert 0\rangle,
\eea
where $\phi = .23\pi$.  Again, all excited crystal field states have been placed at the same energy, here 800K, to reproduce the Van Vleck term,
and again the results are quite insensitive to the exact arrangement of the excited states.}
\item{(C) {\bf Kramers doublet}: $5f^3$ (J = 9/2) $\Gamma_6^{(1)}$:
\begin{equation}
|\Gamma_6^{(1)} \pm\rangle = a\vert \pm 7/2 \rangle +b\vert \mp 1/2\rangle +c \vert \mp 9/2 \rangle,
\end{equation}
where $b$ and $c$ describe the small admixture
of  $|\mp 1/2\rangle$ and $|\mp 9/2 \rangle$ chosen to
 reproduce the anisotropy of the
susceptibility at room temperature.  
We require that we simultaneously 
satisfy the requirement of perfect Ising anisotropy and the correct c-axis moment of 3.5$\mu_B$;
as we discuss in the Appendix, for the Ising state with $5b^2 + 6ac = 0$, this
is only satisfied with a low-lying crystal-field excitation at 110 K that is inconsistent
with current neutron scattering results\cite{Mydosh11} .  
We have chosen a compromise solution to illustrate the 5f$^3$ 
configuration (see Appendix for details) with a crystal-field scheme that is compatible with experiment\cite{Mydosh11}.
This doublet is not a pure Ising doublet, but has a transverse moment 
$\vert\langle-\vert J_+\vert +\rangle\vert =$ $.55\mu_B$, giving a c-axis anisotropy of 1/50.
}
\end{itemize}

Based on
the high-temperature
\emph{linear} susceptibilities (Fig. \ref{fig2}), these three scenarios
are practically indistinguishable. 
The nonlinear susceptibility (Fig 3), $\chi_3^\perp$,
clearly distinguishes the three cases: the non-Kramers doublet has a
clear $+1/T$ dependence, the finely tuned singlets have a $+1/T$ dependence at high temperatures that turns over to become a constant at zero temperature, resulting in a maximum at $T \approx 2\Delta$, where $\Delta$ is the separation of the singlets.  This temperature is significantly higher than the quenching of 
the linear susceptibility, $\chi_1^c$ at $T \approx .25 \Delta$.  By contrast, the Kramers doublet has a strong $-1/T^3$ dependence that should dominate at temperatures much smaller than the crystal field splitting, $T \ll \Delta_{VV}$. 

\fg{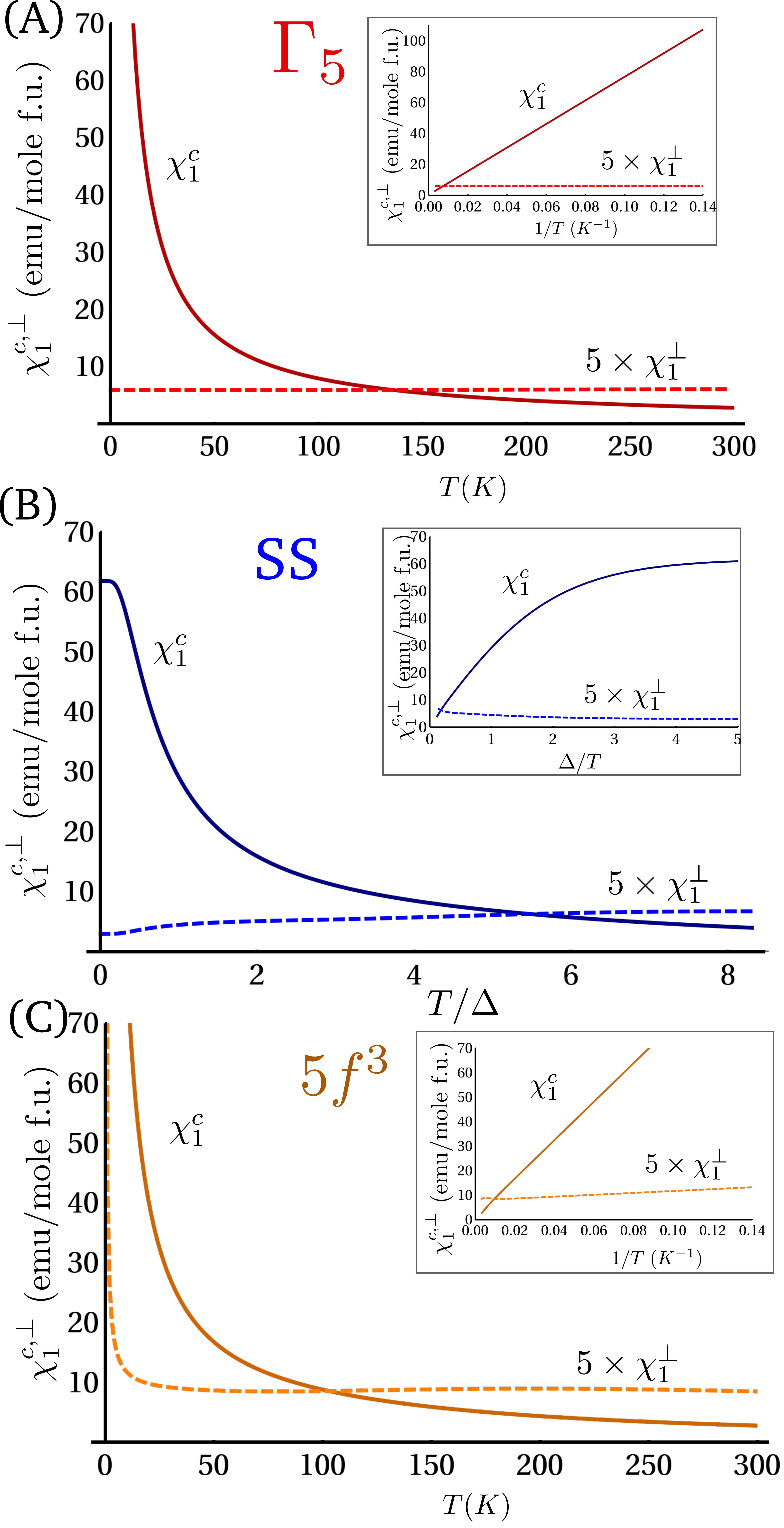}{fig2}{Linear susceptibilities for the three
crystal field schemes with the c-axis and basal plane susceptibilities
given by solid and dashed lines respectively.  The basal-plane
susceptibility is multiplied by five.  (a) $\Gamma_5$
ground-state. Inset shows the anisotropic Curie behavior.
(b) Finely-tuned singlets.  Here the temperature has been rescaled by
the gap, $\Delta = 36K$ between the two singlets. Inset shows the
$1/T$ dependence of the magnetic susceptibility 
at temperatures larger than the gap $\Delta $.
(c) $5f^3$ Kramers
doublet scenario.Inset shows Curie behavior of
susceptibility in both the c-axis and basal plane.}

Of course, these are only single-ion calculations, and in \urs we only expect
single-ion physics to hold for $T > T^* \equiv 70 K$ with ($1/T \rightarrow 1/(T - \theta_{CW}))$,
making it difficult to distinguish the fine-tuned singlets from a true non-Kramers doublet.
In the dilute limit,
U$_x$Th$_{1-x}$Ru$_2$Si$_2$ ($x \leq .07$), the single-ion physics
extends to much lower temperatures, down to 10K.
Below this temperature, the Curie-like single-ion behavior is replaced
by a critical logarithmic temperature dependence, $-\log T/T_K$, where
$T_K \approx 10$K\cite{Amitsuka94}.  This physics has been attributed to two channel
Kondo criticality, although magnetization studies show fine
differences from the single impurity two-channel Kondo model\cite{Toth10} that may
be due to the fact that two-channel impurities are never really in
the idealized dilute limit.

\fg{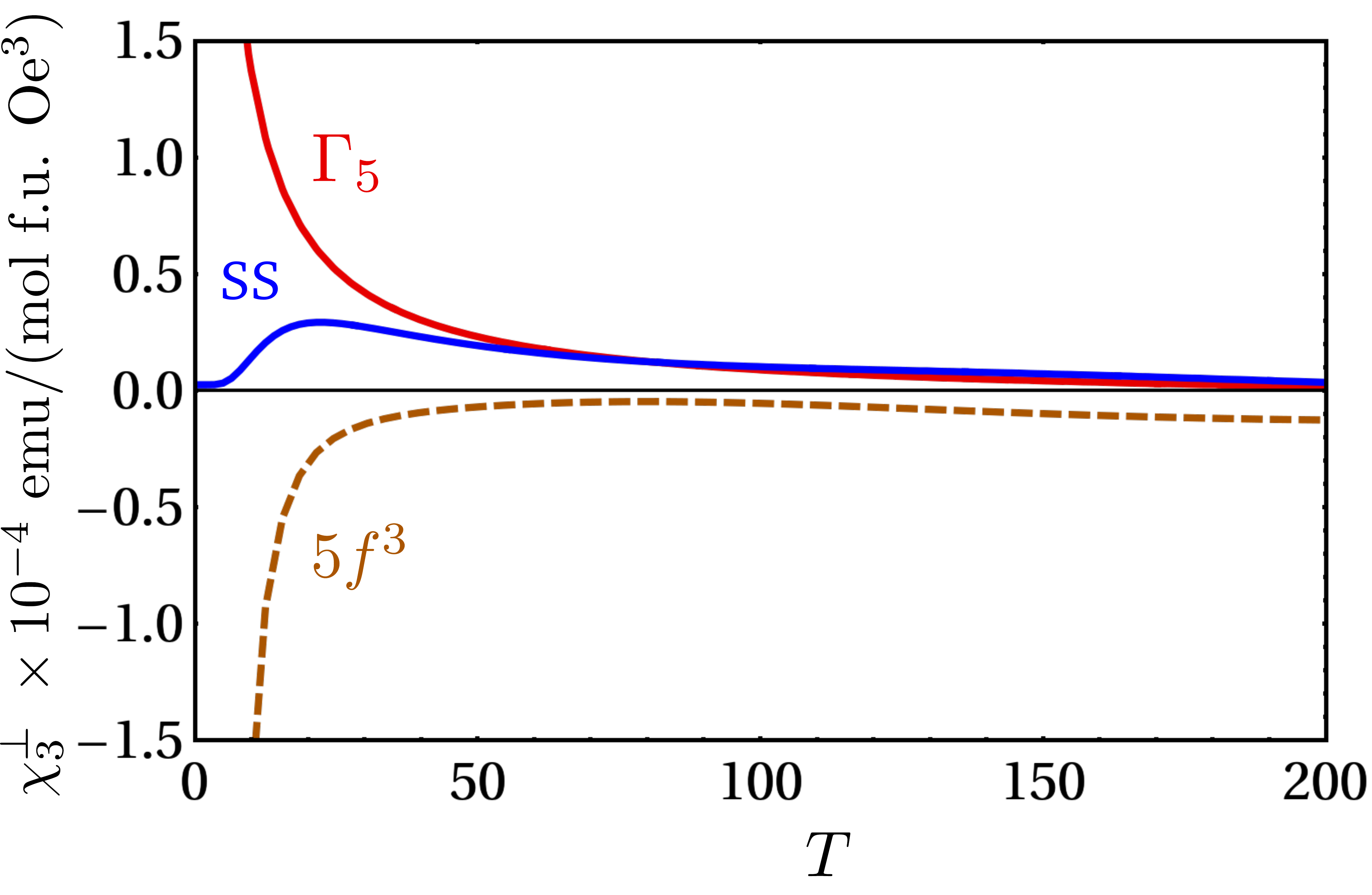}{fig3}
{The basal-plane nonlinear susceptibility, $\chi_3^\perp$ for the three scenarios,
plotted versus the inverse temperature rescaled as $T/\Delta$, where $\Delta = 36 K$
is the gap for the singlet scenario.  The $\Gamma_5$ is given in solid red, the finely-tuned-singlet scenario in solid blue, and the $5f^3$ Kramers doublet scenario in dashed orange  Note that only the singlet scenario has a strong non-monotonic temperature-dependence, with a maximum around $T = \Delta$.  The $\Gamma_5$ scenario has a strong,
positive Curie-dependence, while the $5f^3$ is dominated by a negative 
$-1/T^3$ term.}

\section{IV. Duality of the $\Gamma_5$ doublet}

As $\Gamma_5$ is a magnetic non-Kramers doublet, it is protected by
a combination of both time-reversal and tetragonal
crystal symmetry, ($\vert  \pm \rangle = \vert x \rangle \pm i \vert
y\rangle$), and it thus possesses a dual magnetic and orbital character.
This means that the 
the c-axis moment is magnetic 
(both dipolar and octupolar), while the basal plane moment
is quadrupolar ($\mathcal{O}_{xy}, \mathcal{O}_{x^2-y^2}$).  As magnetic and
quadrupolar moments result in linear and quadratic field splittings, respectively
\begin{equation}
\Delta(B_c,B_\perp) = g_f \mu_F B_c + \frac{1}{2} \alpha \chi_{VV} B_\perp^2,
\end{equation}
they 
will appear
in the linear and nonlinear susceptibilities
respectively, as the nonlinear susceptibility  is essentially the
quadrupolar susceptibility.  Thus the linear c-axis susceptibility and
the nonlinear basal plane susceptibility are predicted to have the same temperature
dependence - albeit with different coefficients.  Taking the above
splitting, the free energy is given by $F = -\frac{\eta}{2} \Delta^2$,
where $\eta = \chi_1^c/(g \mu_B)^2$ is found in c-axis field, $F =
-\frac{1}{2}\chi_1^c B^2_c$.  In a basal plane field,
\begin{equation}
F = -
\frac{\alpha^2 \chi_{VV}^2 \chi_1^c(T)}{8(g \mu_B)^2}B_\perp^4,
\end{equation}
which implies that the basal plane nonlinear susceptibility
\bea
\chi_3^\perp(T) & = & -3\frac{\partial^4 F}{\partial B_\perp^4} = \frac{\alpha^2 \chi_{VV}^2}{(g \mu_B)^2} \chi_1^c(T)\cr
& = & 1.2 \times 10^{-6} \quad  ({\rm T-f.u.})^{-1},
\eea
will always share the temperature dependence of the linear c-axis
susceptibility, where we have used the crystal field parameters for
the $\Gamma_{5}$ doublet scenario (A).  This shared temperature dependence is a result of the
duality of the non-Kramers doublet and will hold not only in the
single-ion limit, but in the critical dilute and dense limits.  
The Kramers doublet does
not have this duality, as we can see from the different temperature
dependences even in the single-ion case. The finely tuned
singlets will only exhibit this duality down to the gap
temperature ($\sim 36K$).

\section{V. Discussion and Conclusions}

Our treatment up to this point has assumed isolated single-ion
behavior, determined by a nominal and integral f-valence. In reality
the 5f state is not a purely integral valence state, but its magnetic properties
do reflect a nominal integral valent configuration.  The reasoning
behind this is somewhat subtle and now deserves discussion.

One of the important concepts that comes to our aid, is the distinction
between the {\sl nominal} integral valence that determines the magnetic
properties and the {\sl microscopic} valence measured by high energy
spectroscopic probes\cite{Anderson83}. To illustrate this point, consider the case
of a nominal $5f^{3}$ configuration, undergoing valence fluctuations
into a $5f^{2}$ singlet, 
\[
5f^{3 } \leftrightharpoons 5f^{2} + e^{-}
\]
In the isolated atomic
limit, such a state is $\vert 5f^{3}:\sigma \rangle $, but once
valence fluctuations are included, this state is then described by a
Varma-Yafet wavefunction\cite{varmayafet} of the form
\begin{equation}\label{}
\vert \sigma^{*}\rangle  = (1 + \sum_{k }\alpha_{k}c\dg_{k \alpha}f_{\alpha })\vert 5f^{3}:\sigma \rangle ,
\end{equation}
where the second term describes the virtual excitation of an
f-electron into the conduction sea in a 
partial wave state with the same spin-orbit coupled 
angular momentum as the f-state.
The important point here, is that
while the valence is reduced below its nominal value, so that here
$n_{f}^{*}= 3-\delta $, since valence fluctuations
conserve total angular momentum, they do not renormalize the magnetic moment.
For this reason, the magnetic moment of an isolated Anderson impurity is
unaffected by significant departures from integral
valence\cite{bickers}.   
The above arguments will of course also apply
if the nominal configuration is a $5f^{2}$ non-Kramers doublet and $n_{f}=2+\delta $.
The important point is that valence fluctuations do not
renormalize the magnetic moment so that a single-ion treatment can be
used to  describe the susceptibility in the region that Curie behavior
is observed. It also means that susceptibility measurements of any
kind can only determine the nominal, not the microscopic valence of
the ion.

There is in fact a lot of circumstantial evidence that these kinds of
arguments apply to \ursp. In particular:
\begin{itemize}
\item  In the dilute limit,  the magnetic susceptibility follows a
Curie Weiss form down to 10$K$, giving us an estimate for the single-ion 
Kondo temperature $T_{K}\sim 10K$. In concentrated case,
single-ion behavior continues down to about 100K, which is still a
small scale more characteristic of a system close to integral valence
than a strongly mixed valent system. 

\item  STM measurements do indicate admixture between the f-electrons
and the conduction electrons. Indeed, coherent tunneling into the
f-states is observed on the Si layers of \urs, indicating that the
renormalized f-states  extend to the silicon orbitals\cite{seamus,yazdani}.

\end{itemize}
Thus  while there is mixed valence, single-ion magnetic physics is observed.

One method to directly determine the microscopic valence of the
$U$ ions is from the branching ratios of two core-level d-states
\cite{Jeffries10}.
The quoted microscopic valence determined by these methods is
$n_{f}=$2.6-2.8, which would suggest a nominal valence of 3 ($5f^{3}$)
with a Kramers doublet. Such measurements would at first sight seem to
make a magnetic determination of the nominal valence unnecessary.
However, recent theoretical work suggests that the microscopic valence
determined by these methods requires a more accurate many body
treatment of the spin-orbit coupling in the f-states, which can give
substantial corrections to the inferred microscopic valence\cite{gabi}.


In conclusion, we have proposed a bench-top experimental probe that
can determine the nominal U valence configuration in \ursp,
distinguishing between a 5f$^{2}$ configuration that is naturally
Ising-like and a 5f$^{3}$ Kramers configuration that is
fine-tuned close to the Ising limit. 
Using
a single ion  approximation valid at high temperatures, 
we have shown the basal plane non-linear susceptibility for
a 5f$^{2}$ non-Kramers or singlet-singlet ground-state is {\sl always
positive}, whereas the Kramers ground state turns negative at low temperatures.
In the dilute limit, we have used the duality of the non-Kramers
doublet to predict a temperature-independent ratio between the basal
plane non-linear susceptibility and the c-axis linear susceptibility;
if found this will provide direct confirmation of the $\Gamma_{5}$
doublet. The nonlinear susceptibility of concentrated \urs was last measured
twenty years ago\cite{Ramirez92,chi3_extra}; measurements were only conducted
for $T < 25 K$, well below the single-ion temperature regime. A set of new measusrements 
over an extended and higher temperature range in both the dilute and the concentrated
limits has the potential to add much insight to this problem. 
Resolution of the uranium single-ion ground state in \urs
will provide crucial insight into the origin of its
hidden order, providing an important constraint on future microscopic
theories.

We acknowledge helpful discussions with E. Abrahams, P.W. Anderson,
N. Butch, K. Haule, G. Kotliar, Y. Matsuda
and J. Mydosh.
This research was supported by funding from the Simons Foundation
(Flint), the National Science Foundation grants NSF DMR 0907179 (Flint,
Coleman) and grant NSF 1066293 (Flint, Chandra, Coleman) while at the Aspen Center for Physics. We are grateful
for the hospitality of the Aspen Center for Physics.

\section{Appendix}

\fg{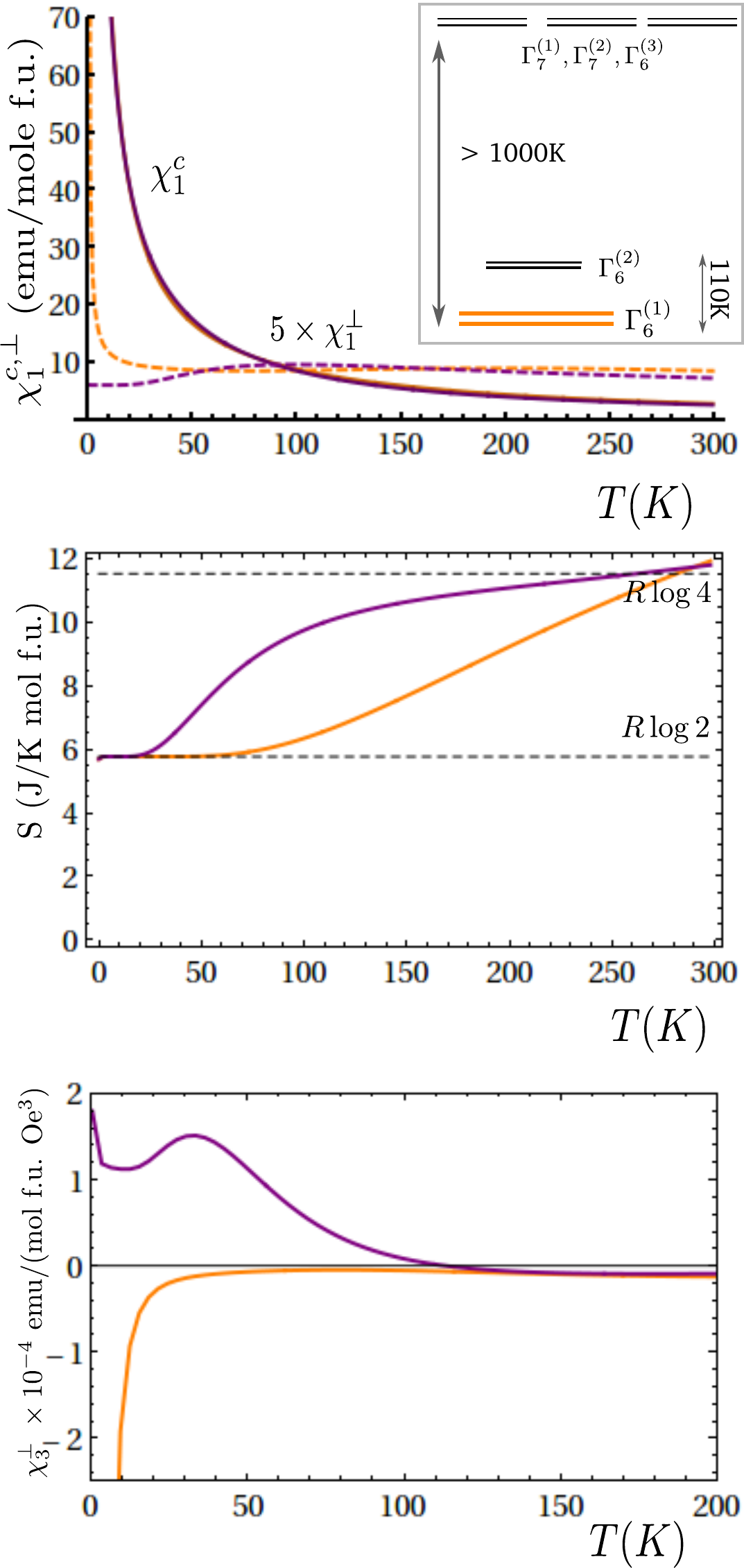}{appendix}{(Color online) Comparing possible 5f$^3$ states: a finely tuned Ising state, $.14 |\pm 9/2\rangle + .39|\pm 1/2\rangle - .91|\mp 7/2\rangle$ (purple, darker) and a more generic compromise state $.05 |\pm 9/2\rangle + .42|\pm 1/2\rangle - .91|\mp 7/2\rangle$ (orange, lighter). (A) The linear susceptibilities both look similar at high temperatures, though the absence of a basal plane moment can be seen at low temperatures. (Inset) Crystal field scheme for the finely-tuned state. (B) The finely-tuned state has a low-lying doublet at 110K that makes the low temperature entropy much larger.  (C) The nonlinear susceptibility looks the same above 100K, but the lack of basal plane moment gives a positive contribution to the finely-tuned state that resembles the $\Gamma_5$ upturn.  However, the negative $\chi_3^\perp(T)$ above 100K clearly indicates the 5f$^3$ nature.}

Here we explain the details of choosing a representative state for the 5f$^3$ configuration.  The strong spin-orbit coupling means that $J=9/2$, and in a tetragonal field, the crystal field Hamiltonian contains five free parameters,
\begin{equation}
H =  B_2^0 O_2^0 + B_4^0 O_4^0 + B_6^0 O_6^0 + B_4^4 O_4^4 + B_6^4 O_6^4,
\end{equation}
where $O_l^m$ are the Stevens operators\cite{Fischer87}.  There are three $\Gamma_6$ doublets that mix $J_z = 9/2$, $7/2$ and $1/2$ states and two $\Gamma_7$ doublets that mix $J_z  = 3/2$ and $5/2$, which between them have eight free parameters, so the problem is grossly overconstrained.  In order to find a set of parameters, $B$ that have the correct c-axis moment, high temperature anisotropy and van Vleck susceptibility, we performed a simulated annealing search of the space of $B$'s.  We found that it is possible to find a state with nearly zero transverse moment, $\vert\langle-\vert J_+\vert +\rangle\vert =$ $.01\mu_B$.  This finely tuned state has a positive $\chi_3^\perp(T)$ at low temperatures due to the lack of basal plane moment.  However, this state also requires the existence of a low-lying doublet at 110K that is inconsistent with neutron scattering\cite{Mydosh11} and gives an entropy of greater than $R\log 4$ at room temperature.  And even in this case, $\chi_3^\perp(T)$ is negative over some regime of temperature, so the sign of $\chi_3^\perp(T)$ is still a good indicator of 5f$^3$ behavior.


\begin{thebibliography}{99}

\bibitem{Mydosh11} J.A. Mydosh and P.M. Oppeneer, P. {\sl Rev. Mod. Phys.} {\bf 83}, 1301-22 (2011).

\bibitem{Palstra85} T.T.M. Palstra et al., {\sl Phys. Rev. Lett} {\bf 55} 2727 (1985).

\bibitem{Miyako91} Y. Miyako et al., {\sl J. Appl. Phys.} {\bf 76}, 5791 (1991).

\bibitem{Ramirez92} A.P. Ramirez et al., {\sl Phys. Rev. Lett.} {\bf 68}, 2680 (1992).

\bibitem{Oppeneer10} P.M. Oppeneer et al, {\sl Phys. Rev. B} {\bf 82}, 205103 (2010).

\bibitem{Broholm91} C. Broholm et al, {\sl Phys. Rev B} {\bf 43}, 12809 (1991).

\bibitem{Park02} J.-G. Park et al., {\sl Phys. Rev. B} {\bf 66}, 094502 (2002).

\bibitem{Jeffries10} J.R. Jeffries et al., {\sl Phys. Rev. B} {\bf 82}, 0334103 (2010).

\bibitem{Severing10}T. Willers et al, {\sl Phys. Rev. B} {\bf 81},
195114, (2010).

\bibitem{Fulde85}P. Fulde and M. Loewenhaupt, {\sl Adv. Phys} {\bf
34}, 589 (1985).

\bibitem{Haule09} K. Haule and G. Kotliar, {\sl Nature Phys.} {\bf 10}, 111 (2009).

\bibitem{Ramirez94} A.P. Ramirez et al., {\sl Phys. Rev. Let.} {\bf 73} 3018 (1994).

\bibitem{Santini94} P. Santini and F. Amoretti, {\sl Phys. Rev. Lett}, {\bf 73}, 1027 (1994).

\bibitem{Sakurai94} J. Sakurai {\sl Modern Quantum Mechanics} (Addison-Wesley, 1994).

\bibitem{Amitsuka94} H. Amitsuka and T. Sakakibara, {\sl J. Phys. Soc. Japan} {\bf 63}, 736-47 (1994).

\bibitem{Toth10}A. Toth, P. Chandra, P. Coleman, G. Kotliar and
H. Amitsuka, {\sl  Phys. Rev. } {\bf 82}, 235116 (2010).

\bibitem{Anderson83} P.W. Anderson in ``Moment Formation in Solids'',
Plenum, ed W. J. L. Buyers, 313-326, (1983).

\bibitem{varmayafet}C. M. Varma and Y. Yafet, Phys. Rev. B 13, 2950 (1976).

\bibitem{bickers}N. E. Bickers, D. L. Cox and J. W. Wilkins, Phys. Rev. B 36, 2036–2079 (1987).

\bibitem{seamus}A. R. Schmidt et al., {\sl Nature} {\bf 465}, 570 (2010).

\bibitem{yazdani}P. Aynajian et al.,  {\sl PNAS} {\bf 107}, 10 383 (2010).

\bibitem{gabi}J. H. Shim, K. Haule and G. Kotliar,
Eur. Phys. Lett. {\bf 87}, 17007 (2009).

\bibitem{Cox87} D.L. Cox, {\sl Phys. Rev. Lett}{\bf 59}, 1240-1243 (1987).






\bibitem{Chandra11} P. Chandra, P. Coleman and R. Flint, Preprint (2011).

\bibitem{Kiss05} A. Kiss and P. Fazekas, {\sl Phys. Rev. B} {\bf 71} 054415 (2005).  

\bibitem{Barzykin95}V. Barzykin and L.P. Gorkov, {\sl Phys. Rev. Lett.} {\bf 74} 4301 (1995).

\bibitem{Fischer87} G. Fischer and A. Herr, Phys. Stat. Sol. (B) {\bf 141}, 580 (1987).

\bibitem{chi3_extra} We note that in Fig. 2. of reference [4], there
is a constant negative $\chi_{3}$ for fields along the a axis. This contribution is
most likely not a single-ion response and may well be  a feature
associated with the heavy electron fluid.  Dilution experiments and
measurements over an extended temperature range will help to resolve
this issue, and in the dense limit, subtracting the ThRu$_2$Si$_2$ nonlinear susceptibility should isolate 
the uranium contribution.

\end{thebibliography}
\end{document}